\documentclass[a4paper, amsfonts, amssymb, amsmath, reprint, showkeys, nofootinbib, twoside,superscriptaddress]{revtex4-1}
\usepackage[english]{babel}
\usepackage[utf8]{inputenc}
\usepackage[colorinlistoftodos, color=green!40, prependcaption]{todonotes}
\usepackage{amsthm}
\usepackage{mathtools}
\usepackage{physics}
\usepackage{xcolor}
\usepackage{graphicx}
\usepackage[left=23mm,right=13mm,top=35mm,columnsep=15pt]{geometry} 
\usepackage{adjustbox}
\usepackage{placeins}
\usepackage[T1]{fontenc}
\usepackage{lipsum}
\usepackage{csquotes}
\usepackage[pdftex, pdftitle={Article}, pdfauthor={Author}]{hyperref} 
\usepackage{mdframed}

\begin{document}
\title{Zooming In on Equity Factor Crowding}

\author{Valerio Volpati}
\affiliation{Capital Fund Management, 23-25, Rue de l'Universit\'e 75007 Paris, France}
\affiliation{Chair of Econophysics and Complex Systems, \'Ecole polytechnique, 91128 Palaiseau Cedex, France}
\author{Michael Benzaquen}
\affiliation{Capital Fund Management, 23-25, Rue de l'Universit\'e 75007 Paris, France}
\affiliation{Chair of Econophysics and Complex Systems, \'Ecole polytechnique, 91128 Palaiseau Cedex, France}
\affiliation{Ladhyx UMR CNRS 7646 and  Department of Economics, \'Ecole polytechnique, 91128 Palaiseau Cedex, France}
\author{Zolt\'an Eisler}
\affiliation{Capital Fund Management, 23-25, Rue de l'Universit\'e 75007 Paris, France}
\affiliation{CFM-Imperial Institute of Quantitative Finance, Department of Mathematics, Imperial College, 180 Queen's Gate, London SW7 2RH}
\author{Iacopo Mastromatteo}
\affiliation{Capital Fund Management, 23-25, Rue de l'Universit\'e 75007 Paris, France}
\affiliation{Chair of Econophysics and Complex Systems, \'Ecole polytechnique, 91128 Palaiseau Cedex, France}
\author{Bence T\'oth}
\affiliation{Capital Fund Management, 23-25, Rue de l'Universit\'e 75007 Paris, France}
\author{Jean-Philippe Bouchaud}
\affiliation{Capital Fund Management, 23-25, Rue de l'Universit\'e 75007 Paris, France}
\affiliation{Chair of Econophysics and Complex Systems, \'Ecole polytechnique, 91128 Palaiseau Cedex, France}
\affiliation{CFM-Imperial Institute of Quantitative Finance, Department of Mathematics, Imperial College, 180 Queen's Gate, London SW7 2RH}
\affiliation{Acad\'emie des Sciences, Quai de Conti, 75006 Paris, France}

\date{\today} 

\begin{abstract}
 Crowding is most likely an important factor in the deterioration of strategy performance, the increase of trading costs and the development of systemic risk.  We study the imprints of \emph{crowding} on both anonymous market data and a large database of metaorders from institutional investors in the U.S. equity market.
We propose direct metrics of crowding that capture the presence of investors contemporaneously trading the same stock in the same direction by looking at fluctuations of the imbalances of trades executed on the market. We identify significant signs of crowding in well known equity signals, such as Fama-French factors and especially Momentum. We show that the rebalancing of a Momentum portfolio can explain between 1--2\% of order flow, and that this percentage has been significantly increasing in recent years.
\end{abstract}

\keywords{crowding, equity factors, momentum, market microstructure}

\maketitle


\section{Introduction} 

``Crowded trades'' or ``crowded strategies'' are often heard explanations for the sub-par performance of an investment or, in more extreme cases, for the occurence of 
deleveraging spirals. Although seemingly intuitive, the concept of crowding has remained elusive and is, in fact, somewhat paradoxical as every buy trade is executed against a sell trade of the same magnitude. Some clarification is therefore needed, and the subject has recently garnered substantial interest in academic \cite{deleveragingspirals,khandani2007happened,overlapping,overlapping2,barroso2019institutional} and applied research \cite{CFMsardines}. 

Investors in a purportedly crowded strategy may face three related predicaments. One is that of increased competition for the same excess returns, leading to an erosion of the performance of the strategy. Second is increased transaction costs: maintaining similar portfolios leads to similar trade flows. This amplifies the effective market impact suffered by all investors following the same strategy -- an effect called co-impact in \cite{bucci2018co}. This in turn leads to a deterioration of performance even under normal conditions, see e.g. \cite{CFMimpact}.  Finally, if the portfolios of different competitors largely overlap, systemic risk may arise as the liquidation of one of these portfolios can trigger further liquidations and even severe cascading losses for all investors who shared similar positions \cite{deleveragingspirals,overlapping,barroso2019institutional}. This phenomenon is well exemplified by the Quant Crunch of 2007 \cite{khandani2007happened}, which chiefly affected a certain style of relative value investing, while it left the market index itself, and therefore long-only investors, largely unscathed.  

Crowding has recently been invoked in the context of Equity Factor strategies, which have witnessed substantial inflows in the past decade, and are currently (as of end 2019) in a relatively severe drawdown. These strategies are based on persistent anomalies which are well known to investors, such as Momentum, Value, or Small Cap strategies \cite{fama1992cross}. This makes them potentially crowded and thus interesting to investigate. 

The aim of the present study is to investigate the possible crowdedness of Equity Factor strategies by measuring the correlation of market order flow with the strength of the trading signal that factor investors hypothetically follow. We use both (i) the total order flow measured using anonymized microstructure data pertaining to stocks of the Russell 3000 index and (ii) the institutional order flow identified thanks to a proprietary database. Although these correlations are small, $\sim 1\%$, they are strongly significant and are seen to have increased over the recent years. The estimated impact costs suggest that simple Fama-French factor investing is close to saturation.  

\section{Data and Metrics}

This paper relies on three different datasets: one using equity prices to construct Equity Factor trading signals, and two allowing us to quantify trading activity.

\subsection{Equity Factor Data}

\begin{figure}[!htb]
  \includegraphics[width=\columnwidth] {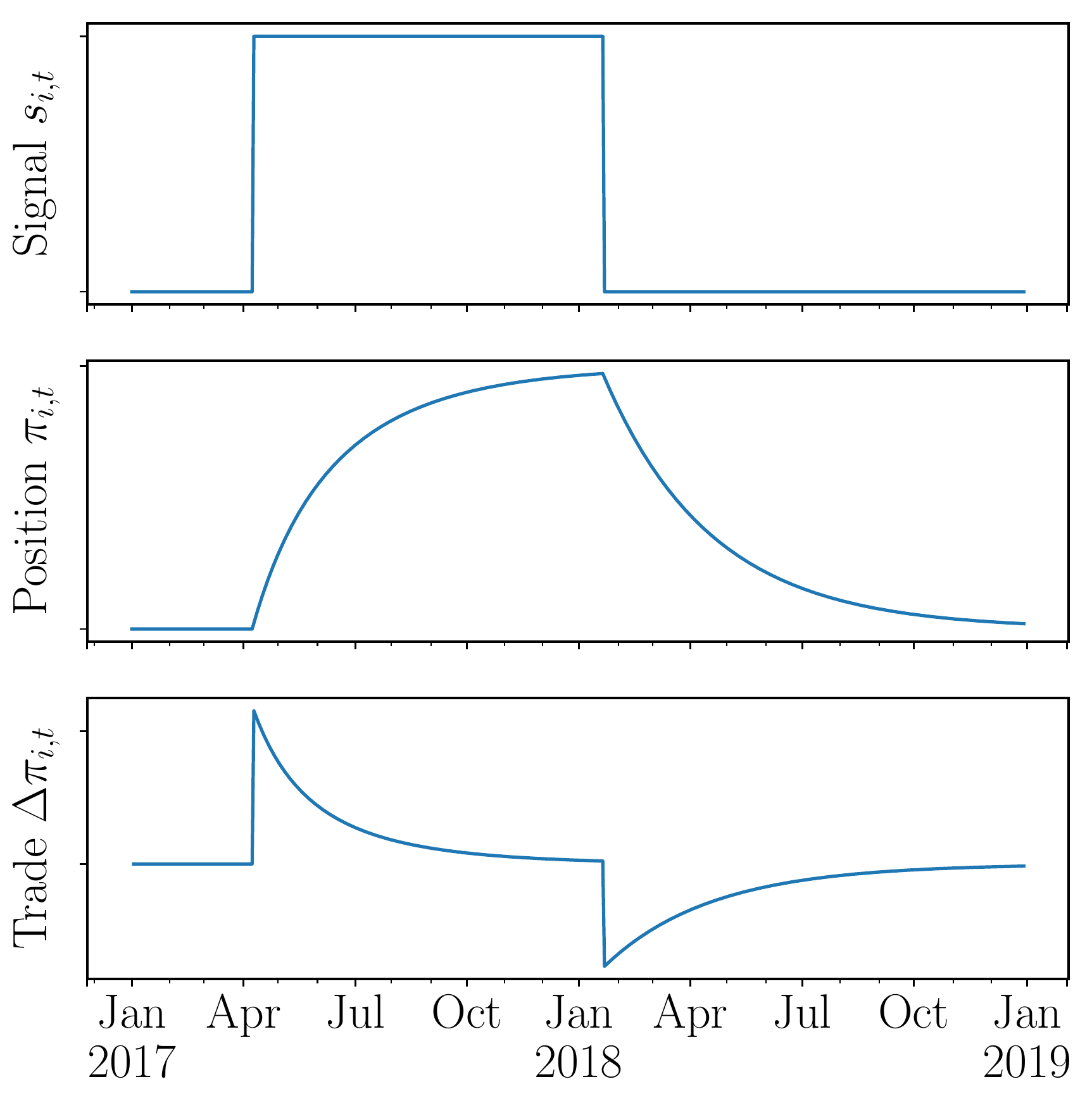}
  \caption{\textbf{The slowing down procedure implemented in this paper}. (Top panel) The original signal $s_{i,t}$, corresponding to ideal position before trading costs are considered. 
  (Middle panel) The slowed down signal $\pi_{i, t}$ with slowing down timescale $D=3$ months, corresponding to the desired position when transaction costs are taken into account.
  (Bottom panel) The expected order flow $\Delta \pi_{i, t}$, submitted by the investor who is targeting the position $\pi_{i, t}$.
  }
  \label{fig:signalsketch}
\end{figure}

We use standard Fama-French (FF) factors \cite{fama1992cross,ang2014}, extended to Momentum \cite{jegadeesh1993returns, carhart1997persistence}, defined on the components of the Russell 3000 index in a period spanning from January 1995 to December 2018. Since rebalancing these FF portfolios is costly, we expect investors to slow down the bare signal to trade less aggressively. A conceptually sound \cite{garleanu2013dynamic} model that assumes quadratic trading costs leads to an exponential slowing down of the signal.  More formally, let us denote by $s_{i,t}$ the ``signal'' followed by an investor, giving the ideal holding of stock $i$ on the close of day $t$. For example, $s_{i,t}$ would be the ranking of stocks according to their past returns in the case of the Momentum factor. The {\it actual} holdings  $\pi_{i, t}$ of the investor is then given by an Exponential Moving Average (EMA):
\begin{equation}
    \pi_{i, t} =  A \, \sum_{t' \leq t} s_{i, t'} \,\, \exp\left(-\frac{t-t'}{D}\right).
\end{equation}
The factor $A$ sets the overall risk of the portfolio, whereas the slowing down time scale $D$ is chosen as to provide a good compromise between performance and trading costs. The theoretically expected order flow from the strategy on day $t$ is thus given by
\begin{equation}
    \Delta \pi_{i, t} = \pi_{i, t}-\pi_{i, t-1}.    
\end{equation}
This framework is summarized in Figure \ref{fig:signalsketch}.

\subsection{Market Microstructure Data}

We also use anonymous market data collected by Capital Fund Management (CFM) covering about $1,600$ US stocks between January 2011 to May 2018. We process a large majority of trades executed throughout different market venues. For each stock $i$ and each day $t$, we define the {\it trade imbalance} as
\begin{equation}
I^\mathrm{trade}_{i, t} = \frac{\sum_{n \in t} \epsilon_{i,n}}{\sum_{n \in t} |\epsilon_{i,n}|},
\end{equation}
where $\epsilon_{i, n}$ is the sign of the $n$'th trade on stock $i$, and the sum extends to all trades taking place during the continuous trading session of day $t$. The sign is considered positive if the trade was executed above the prevailing mid price, and negative otherwise. Midprice trades are excluded. Trades are generated by aggressive orders; $I^\mathrm{trade}_{i,t}$ hence captures the pressure that they imply. 

Similarly, we define the {\it volume imbalance} as
\begin{equation}
 I^\mathrm{volume}_{i, t} = \frac{\sum_{n \in t} \epsilon_{i,n} v_{i,n}}{\sum_{n \in t}  v_{i,n}},
\end{equation}
where $v_{i,n}$ is the volume executed on stock $i$ at trade $n$. We do not expect, and in fact do not find, major differences between the behavior of $I^\mathrm{trade}$ and $I^\mathrm{volume}$, because volumes are constrained by liquidity available at the best quotes, and do not fluctuate wildly.

To capture the liquidity in the order book, we average the volume available on the bid side $V^\mathrm{bid}_{i, s}$ and the one available on the ask side $V^\mathrm{ask}_{i, s}$  over $N$ snapshots $s$ taken every $5$ seconds during the continuous trading session of day $t$, to calculate, for a given day and stock
\begin{equation}
V^\mathrm{bid}_{i, t} = \frac{1}{N} {\sum_{s\in t} V^\mathrm{bid}_{i, s}},
\end{equation}
and
\begin{equation}
V^\mathrm{ask}_{i, t} = \frac{1}{N} {\sum_{s\in t} V^\mathrm{ask}_{i, s}}.
\end{equation}
These allow us to define the daily average \emph{order book imbalance} as
\begin{equation}
I^\mathrm{book}_{i, t} = \frac{V^\mathrm{bid}_{i, t} - V^\mathrm{ask}_{i, t}}{V^\mathrm{bid}_{i, t} + V^\mathrm{ask}_{i, t}} .
\end{equation}

All imbalances above are, by definition, bounded between -1 and 1.

\subsection{Metaorder Data}

Finally, we also use the Ancerno dataset, a proprietary database containing trades of institutional investors, covering about $10,000$ execution tickets per day, which corresponds to approximately $10\%$ of the daily volume traded on the market from January 1999 to December 2014. This data set has been used in several academic studies in the past, see e.g. \cite{ganghu2018,busse2017,zarinelli,bucci2018co,bucci2019crossover}, and we refer to those papers for more information on the data set, as well as descriptive statistics. 
Using labels present in the data, we are able to group together different trades that belong to the same metaorder, i.e. an ensemble of trades with the same client number, start date, end date, stock symbol, and sign. This allows us to gain more information about the decision-making process underpinning the order flow. Furthermore, Ancerno trades are representative of institutional investors, for whom we expect a larger propensity to follow classical equity factors. 

Similarly to the above measures of imbalance, we can define a \emph{metaorder trade imbalance} $I^\mathrm{meta}_{i, t}$ and a \emph{metaorder volume imbalance} $I^\mathrm{metavolume}_{i, t}$ by restricting the sums over all market trades to metaorders executed on stock $i$ on day $t$. For example:
\begin{equation}
I^\mathrm{meta}_{i, t} = \frac
{\sum_{m \in t} \epsilon^\mathrm{meta}_{i,m}}
{\sum_{m \in t} |\epsilon^\mathrm{meta}_{i,m}|}
\end{equation}
where $\epsilon^\mathrm{meta}_{i,m}$ is the sign of metaorder $m$ on stock $i$, and where $m$ runs on the total number of identified metaorders on stock $i$ and day $t$. 

\section{Return-Imbalance Correlations}

Before diving into the original part of our study, namely the correlation between imbalances and trading signal, we re-establish some well known facts about imbalance-return and imbalance-imbalance correlations. 

\begin{figure}[!htb]
  \includegraphics[width=0.5\textwidth] {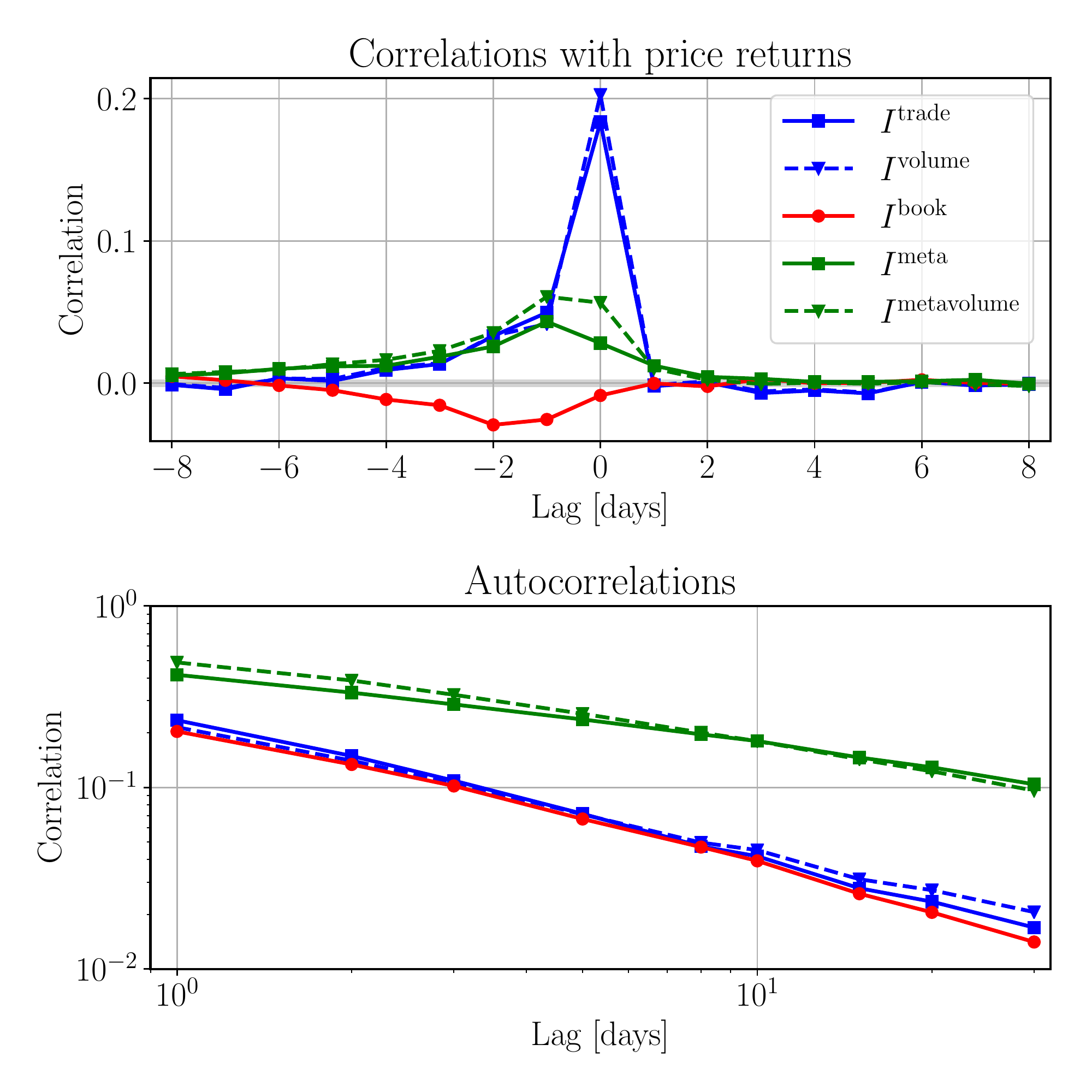}
  \caption{ \textbf{Correlation of imbalances with price returns and autocorrelations}.
  (Top panel) Lagged correlations of all imbalances at day $t$, with price returns at day $t+\textrm{Lag}$. For $\textrm{Lag}>0$ imbalances built with public information have no correlation with future returns as expected. For $\textrm{Lag}=0$ we recover the well known ``mechanical'' correlation between contemporaneous trade imbalance and returns. For $\textrm{Lag}<0$ we observe how today's imbalances are correlated to past returns.
  (Bottom panel) We show the time autocorrelation of all imbalances introduced in the text. While price returns are only weakly autocorrelated, the submission of orders, and in particular metaorders, are strongly autocorrelated, with a power-law decay $\text{Lag}^{-\gamma}$ of the autocorrelation function (see \cite{slowlydigest2009, bouchaud2018trades}). We find $\gamma \approx 0.8$ for market-wide trades and $\gamma \approx 0.5$ for metaorders. 
  } 
  \label{fig:correls2}
\end{figure}

Figure \ref{fig:correls2}, top panel, shows the lagged correlation between different imbalance measures and returns, averaged over all the stocks in our dataset. For zero lag, we recover the usual positive correlations between trade imbalance and returns, here found to be $\approx 0.2$. For positive lags, i.e.~when returns are posterior to imbalances, correlation is close to zero, indicating that past imbalances have on average no linear predictability on future returns. On the other hand, past returns are found to be positively correlated with future trade imbalance, both for the market-wide $I^\mathrm{trade}$ and for the metaorder imbalance $I^\mathrm{meta}$. This is compatible with the strong temporal autocorrelation of trade imbalance, documented in \cite{slowlydigest2009, bouchaud2018trades}, and again shown in Figure \ref{fig:correls2}, bottom panel.

We also mention that we find a clear positive correlation between metaorder imbalances $I^\mathrm{meta}$ and market-wide imbalances $I^\mathrm{trade}$ (not shown). This has to be the case since metaorders themselves contribute to the anonymous market flow. We also find a negative correlation of a few percent between metaorder imbalances and book imbalances. This can be understood by arguing that metaorders are more likely to be executed when there is enough available liquidity, and vice-versa: the execution of large metaorders induces more liquidity to be revealed in the order book (a phenomenon called liquidity refill in \cite{bouchaud2018trades}). 

\section{Dynamical Correlations}

\subsection{Momentum}

Let us now turn to the study of correlations between different kinds of imbalances and factor trading, starting with the standard Momentum Factor \cite{jegadeesh1993returns,carhart1997persistence} for which we see the most significant results. As explained above, we first slow down the signal using an EMA, and then take the derivative to calculate the trades that would follow from rebalancing.

\begin{figure}[!htb]
  \includegraphics[width=0.5\textwidth] {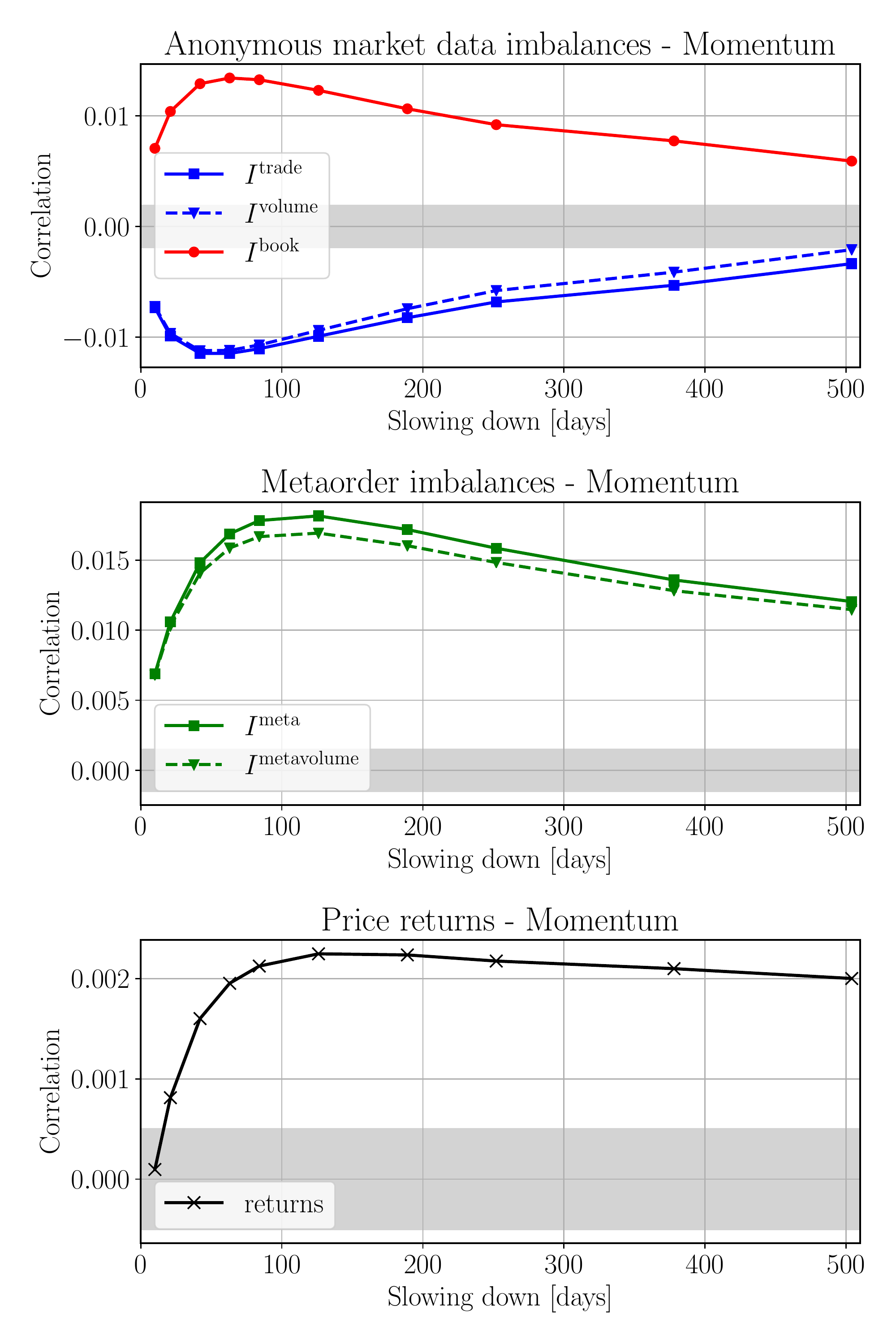}
  \caption{ \textbf{Conditional correlations of all imbalances with slowed down momentum.} (Top panel) Average correlation of market-wide trade imbalances and of book imbalance, with the Momentum trading signal slowed down with different timescales $D$, on the $x$ axis. (Middle panel) Average correlation between metaorder imbalances and the Momentum trading signal, slowed down with different timescales $D$ (Bottom panel) Average correlations between the daily close to close price returns and the Momentum trading signal, slowed down with different timescales $D$. The latter correlation displays a qualitatively similar behaviour, but it is 10 times smaller ($\sim 0.2 \%$) than the correlation between imbalances and trading signal.}
  \label{fig:resmomentum}
\end{figure}

In the top panel of Figure \ref{fig:resmomentum} we show the correlation of anonymous market imbalances with the expected Momentum order flow for different values of the slowing down parameter $D$, averaged over all stocks in the dataset. 
The grey stripe denotes the significance band of the correlations, obtained by reshuffling the time series (in blocks of 6 months in order to preserve the autocorrelation structure) and calculating the standard deviation of the obtained correlations over 200 reshuffled samples.
We find a significant correlation between trade and book imbalances and the expected order flow. This correlation is {\it negative} for trade imbalance and {\it positive} for order book imbalance, with absolute maxima $\gtrsim 1\%$ for $D$ around 3--4 months. This timescale is of the same order of magnitude as the autocorrelation time of the momentum signal, and thus quite reasonable. 

Although the results in the top panel of Figure \ref{fig:resmomentum} are highly significant, the sign of the correlations is somewhat confusing. Naively, one would be tempted to argue that Momentum investors execute their trades with aggressive market orders. This should lead to a positive correlation between the trading signal and trade imbalance, whereas we observe this correlation to be negative. At this stage, one can come up with two opposite interpretations:
\begin{itemize}
    \item What we see are in fact aggressive orders by mean-reversion traders. However, this is quite unlikely since mean-reversion is not a profitable strategy on the time scale of months, but rather on the time scale of a few days only.
    \item Momentum trades are predominantly executed using limit orders, thus contributing to a positive correlation with the order book imbalance, as observed in the top panel of Figure \ref{fig:resmomentum}. This is compatible with the behavior of popular broker execution algorithms that chiefly use passive orders. It also resonates with \cite{aqr_liquidity}, which states that 80 \% of the volume executed by a major fund manager in the factor trading industry is through limit orders. 
\end{itemize}
The analysis of metaorders helps bolster our interpretation. In the middle panel of Figure \ref{fig:resmomentum} we see that both metaorder and metaorder volume imbalance show a clear {\it positive} correlation with the Momentum trading signal, with a similar time scale $D$ around 4--6 months. Note that the sign of metaorders (to buy or to sell) reported by the data provider is not sensitive to whether they are executed actively or passively. Since most metaorders in the database correspond to order flow of institutionals who most likely engage in Momentum strategies, we can safely conclude that the inverted correlation observed in the top panel of Figure \ref{fig:resmomentum} corresponds to Momentum trading using passive orders. 

We have performed a series of tests for the robustness of the observed correlations. These are quite stable across stocks, regardless of liquidity and tick size. The same analysis with different slowing down mechanisms, or using the long (or short) only component of Momentum yields qualitatively similar results in all cases.

\begin{figure*}[!ht]
  \includegraphics[width=\textwidth] {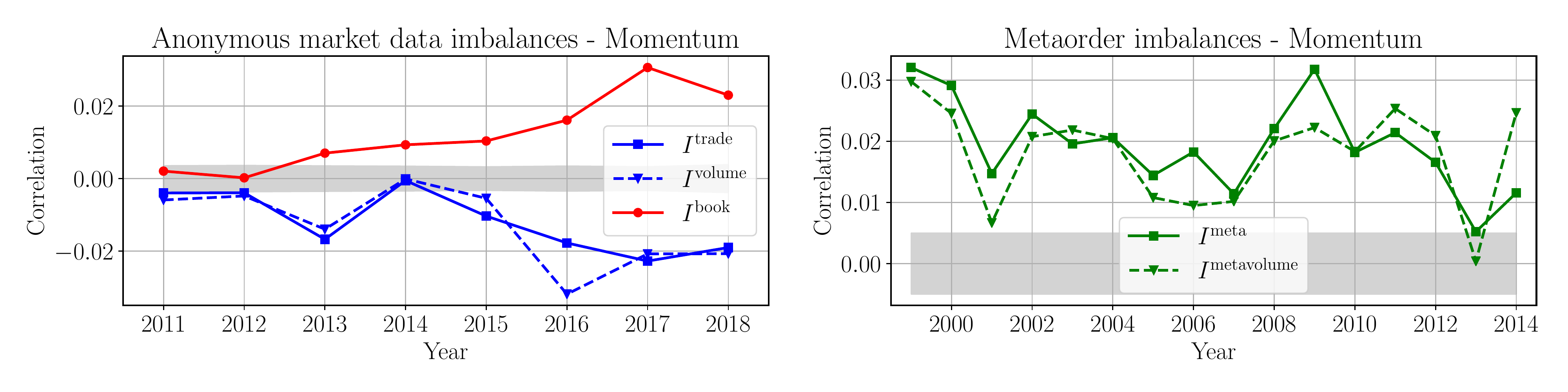}
  \caption{\textbf{Time evolution of the correlation between imbalances and Momentum signal.} (Left panel) Maximum (in absolute value) of the trade imbalance and book imbalance correlation with the Momentum trade signal versus time, since 2011. One observes a clear upward trend, suggesting that Momentum trading has become more and more crowded. (Right panel)  Maximum (in absolute value) of the metaorder imbalance correlation with the Momentum trade signal versus time, from 2000 to 2014. The sign and absolute values of these correlations, albeit noisy, are quite stable in time.}
  \label{fig:resmomentum2}
\end{figure*}

We also show the correlations between the daily close-to-close price returns and the Momentum trading signal, which we can compute on the Russell 3000 from 1996 to 2019, i.e. with much more data than for imbalance correlations. This correlation is considerably smaller: its maximum is around $0.2\%$, to be compared with $1.2\%$ for the imbalance correlation.\footnote{A simple, back-of-the-envelope estimation based on a linear impact model allows one to explain the observed factor $5$ between return/signal and imbalance/signal correlations, taking into account the fact that the unconditional standard deviation of $I^\mathrm{volume}$ is $\sim 0.1$. However, one would rather expect a square-root impact and not a linear impact \cite{bucci2018co}, so it is at this stage not clear how to reconcile these observations.}  Although quite small, this correlation is important as it can be used to estimate the impact cost incurred by Momentum traders. It should be compared with the correlation between the slowed down position $\pi_t$ and the returns, which by definition gives the average profit of the strategy, and is found to be $\approx 0.1 \%$. Assuming a quadratic cost model gives a trading cost equal to one half of the instantaneous impact. These numbers therefore suggest that Momentum has become only marginally profitable: this is a crowded trade, for which co-impact has driven profits to zero. Let us stress that this is not the case for other implementations of Momentum that are, according to the same measure, distinctively less crowded and therefore still profitable (at least at the time of writing). 

We can in fact prove directly that Momentum crowding has increased in the recent years, by computing the maximal imbalance/signal correlation computed every year, found for $D \in (2,4)$ months for market-wide data and $D \in (3,6)$ months for metaorder data. The results are shown in Figure  \ref{fig:resmomentum2} and are quite consistent over the whole period. Whereas the maximum correlation is noisy but quite stable for metaorders from 1999 to 2014, market-wide data that we collect up to 2018 reveals a clear upward drift since 2012, possibly related to the increase of the popularity of factor strategies. 

\subsection{Other Factors: HML \& SMB}

\begin{figure}[!htb]
  \includegraphics[width=0.5\textwidth] {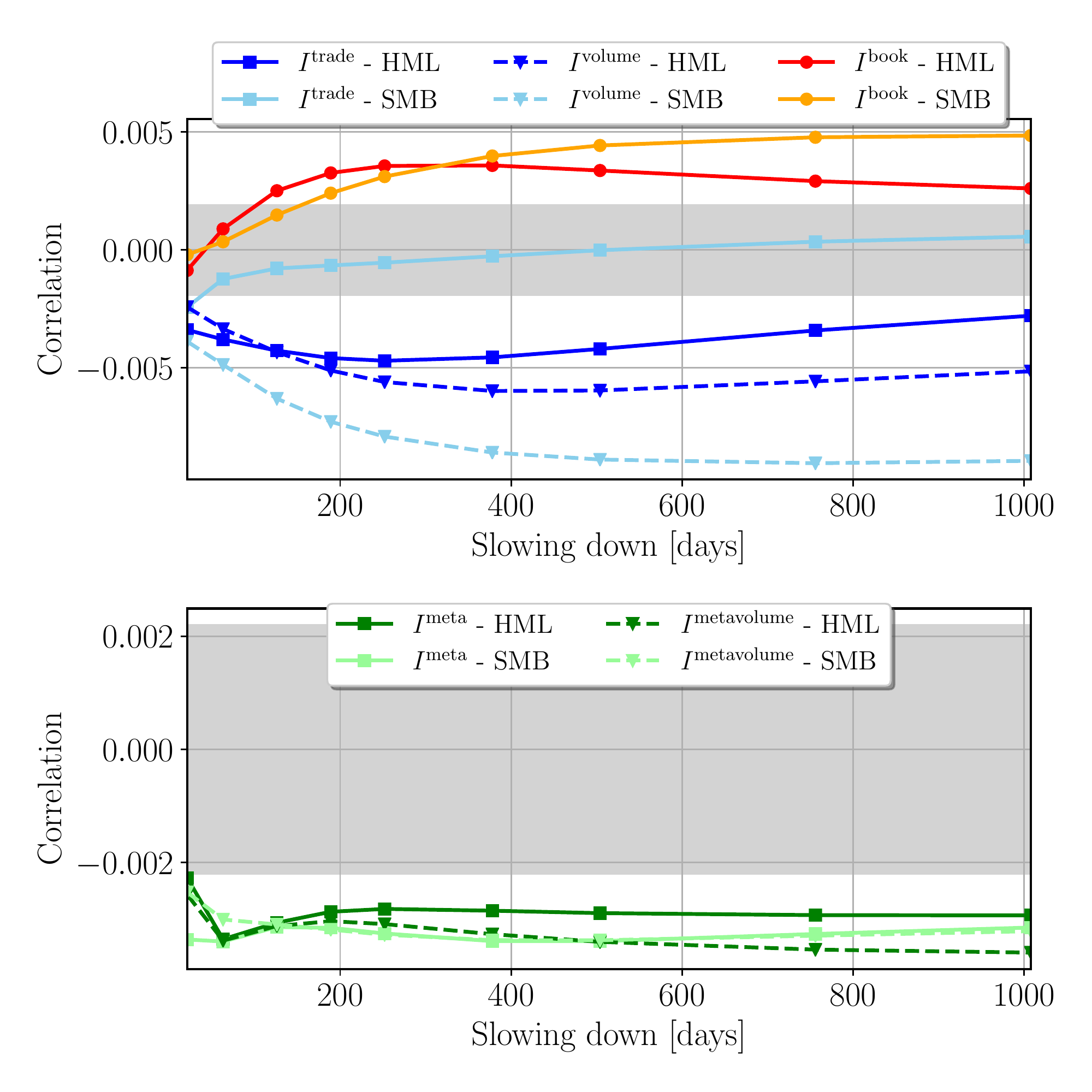}
  \caption{ \textbf{Conditional correlations of imbalances with slowed-down HML and SMB}. We again show, as a function of $D$, the correlation between different types of imbalances and the trading signal originating from the HML and SMB factors. Because these factors have a significantly longer intrinsic time scale, the signal is weaker and shifted to larger values of $D$. Note that the correlation of the trading signals to metaorder imbalances is barely significant, and points in the opposite direction.}
  \label{fig:otherfff}
\end{figure}

So far we discussed results for Momentum only. We also explored Fama-French factors, such as ``HML'' (also called ``Value''), comparing the price of a stock to its book value, and ``SMB'', i.e. buying small cap stocks and selling large cap stocks \cite{ang2014}. Our methodology closely follows that of the previous section. In Figure \ref{fig:otherfff} we show our results for HML and SMB.  Our results for market-wide imbalances for these two factors are similar to those obtained for momentum --~although less significant for SMB. This is expected, given that the longer holding period of these strategies induces a much smaller rebalancing activity~\cite{briere2019stock}. For metaorder imbalance, on the other hand, correlations are barely significant. 

On the other hand, the time evolution of the market-wide imbalances with the HML or SMB signals is too noisy to confirm that crowdedness in these factors has also increased in the recent years.  

\section{Conclusions}

In this empirical study we have shown that crowding of equity factors can be quantitatively elicited through correlations between real supply/demand imbalances (proxies of market participants trading in the same direction) and the rebalancing order flow that would result from tracking slowed-down equity factors. Our results, particularly significant for Momentum, show that such a strategy is indeed crowded, resulting in rather poor profitability. Further, our method allows one to confirm that crowding on equity factors (at least on Momentum) has, as claimed and feared by many, significantly increased over the past few years.

\section{Acknowledgments}
The authors thank Charles-Albert Lehalle for his initial suggestion to study order flow in relation to Ancerno metaorders. We would also like to thank Stefano Ciliberti, Philip Seager and Juha Suorsa for interesting discussions on the subject. We would also like to acknowledge the valuable help and suggestions of Matthieu Cristelli. This research was conducted within the \emph{Econophysics \& Complex Systems} Research Chair, under the aegis of the Fondation du Risque, the Fondation de l’Ecole polytechnique, the Ecole polytechnique and Capital Fund Management.

\section*{Data availability statement}
The data were purchased by Imperial College from {AN}cerno Ltd (formerly the Abel Noser Corporation) which is a widely recognised consulting firm that works with institutional investors to monitor their equity trading costs. Its clients include many pension funds and asset managers. The authors do not have permission to redistribute them, even in aggregate form. Requests for this commercial dataset can be addressed directly to the data vendor. See www.ancerno.com for details.

\bibliography{biblio}{}
\bibliographystyle{abbrvnat}

\end{document}